\documentclass[12pt]{article}

\usepackage{a4wide}
\usepackage{color}

\usepackage{graphicx}

\usepackage{amsfonts}
\usepackage{amsmath}
\usepackage{amssymb}
\usepackage{theorem}
\usepackage{booktabs}
\usepackage{rotating}
\usepackage{nicefrac}
\usepackage[colorlinks,linkcolor=red,anchorcolor=blue,citecolor=blue]{hyperref}
\allowdisplaybreaks

\newtheorem{example}{Example}

\catcode`@=11 \@addtoreset{equation}{section} \catcode`@=12

\numberwithin{equation}{section}
\newcommand{\ba}{\begin{equation}}
\newcommand{\ea}{\end{equation}}

\newcommand{\skellam}{\textup{Sk}}
\newcommand{\poi}{\textup{Poi}}
\newcommand{\bin}{\textup{Bin}}

\newcommand{\norm}{\textup{N}}

\newcommand{\bbn}{\mathbb{N}}
\newcommand{\bbr}{\mathbb{R}}
\newcommand{\bbz}{\mathbb{Z}}

\newcommand{\ie}{i.\,e., }
\newcommand{\eg}{e.\,g., }

\usepackage{dsfont}
\newcommand{\indfkt}{\mathds{1}}

\usepackage{natbib}

\usepackage{hyperref}

\begin{document}




\title{The Mollified (Discrete) Uniform Distribution and its Applications}
\author{
Christian H.\ Wei\ss{}\thanks{
Helmut Schmidt University, Department of Mathematics and Statistics, Hamburg, Germany.}\ \thanks{Corresponding author. E-Mail: \href{mailto:weissc@hsu-hh.de}{\nolinkurl{weissc@hsu-hh.de}}. ORCID: \href{https://orcid.org/0000-0001-8739-6631}{\nolinkurl{0000-0001-8739-6631}}.}
}

\maketitle

\begin{abstract}
\noindent
The mollified uniform distribution is rediscovered, which constitutes a ``soft'' version of the continuous uniform distribution. Important stochastic properties are presented and used to demonstrate potential fields of applications. For example, it constitutes a model covering  platykurtic, mesokurtic and leptokurtic shapes. Its cumulative distribution function may also serve as the soft-clipping response function for defining generalized linear models with approximately linear dependence. Furthermore, it might be considered for teaching, as an appealing example for the convolution of random variables. Finally, a discrete type of mollified uniform distribution is briefly discussed as well.

\end{abstract}

\section{Introduction}
\label{Introduction}
This article deals with a surprisingly little-known distribution, namely the \emph{mollified uniform distribution}, the cumulative distribution function (cdf) of which is also known as the \emph{soft clipping function}. It has been occasionally used in practice, \eg as an output or activation function for artificial neural networks (ANNs) like in \citet{klimek20}, or as a response function for time series models like in \citet{weissjahn24}, and it has been discussed in forums such as ``Cross Validated'' (\url{https://stats.stackexchange.com/q/332558}, accessed on 18 November 2024). 
But there is hardly any literature about the distribution itself and its potential applications. The low level of awareness is also surprising in view of the fact that the mollified uniform distribution would be well suited for teaching, namely as a non-trivial (and quite appealing) example for the convolution of random variables. Accordingly, this article intends to dispel the aforementioned shortcomings by providing a concise review about the different facets of the mollified uniform distribution and its possible applications. 

\begin{figure}[t]
\center\scriptsize
	(a)\hspace{-4ex}\includegraphics[viewport=0 45 300 305, clip=, scale=0.45]{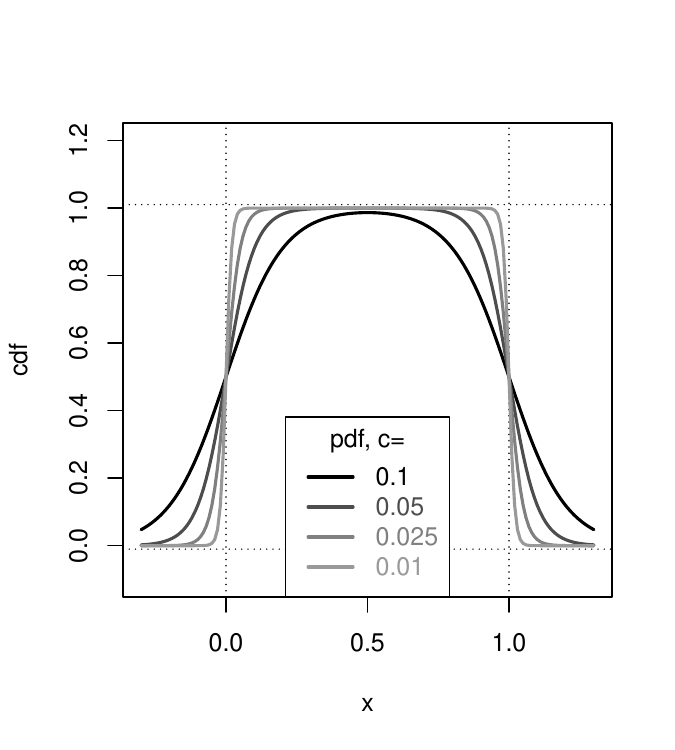}$x$
	\quad
	(b)\hspace{-4ex}\includegraphics[viewport=0 45 300 305, clip=, scale=0.45]{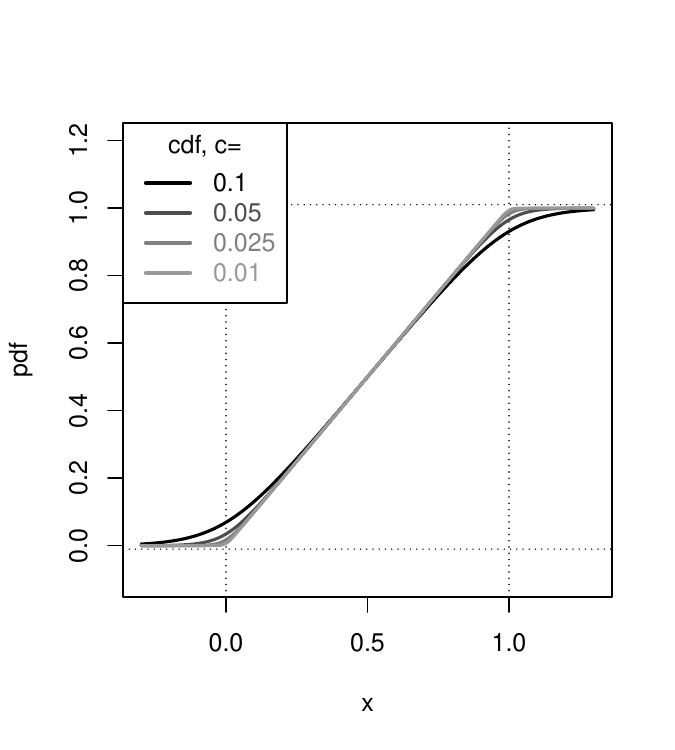}$x$
	\quad
	(c)\hspace{-4ex}\includegraphics[viewport=0 45 300 305, clip=, scale=0.45]{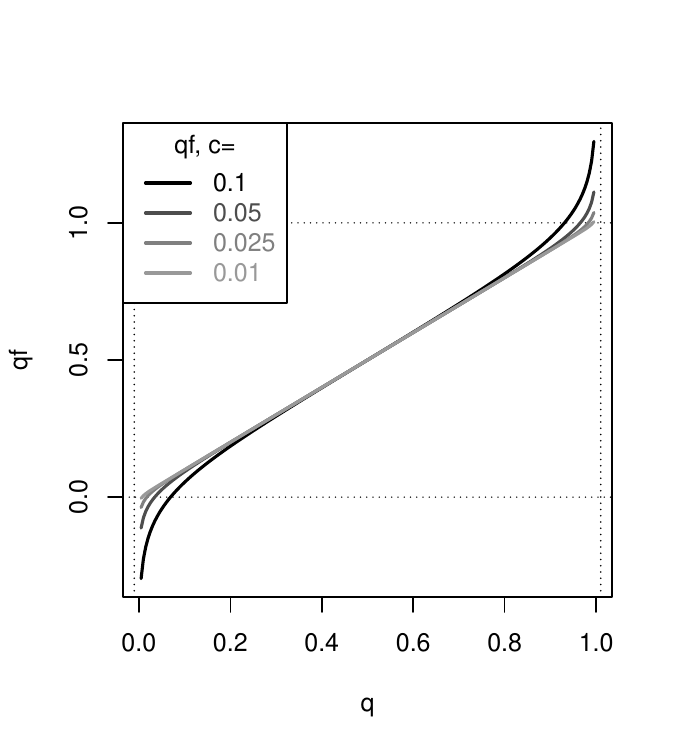}\,$q$
	\caption{Logistic-mollified uniform distribution with scale parameter~$c$: plots of (a) pdf, (b) cdf, and (c) qf for different~$c$.}
	\label{figmollified}
\end{figure}

\smallskip
The idea of mollifying the (standard) uniform distribution $\textup{U}(0,1)$ dates back to \citet{friedrichs44}, who proposed ``a certain class of smoothing operators approximating unity'' (p.~132), namely by adding an independent and centered random variable~$L$ (the mollifier) with very low dispersion to the uniform random variable~$U$ (for properties of the uniform distribution, see \citet[Ch.~26]{johnson95}). While \citet{friedrichs44} originally focused on mollifiers~$L$ with compact support, it is more convenient to consider random variables~$L$ on the full set~$\bbr$ of reals, having a symmetric distribution around zero and arbitrarily low dispersion. These could be a normally distributed $L\sim\norm(0,\sigma)$ with (low) standard deviation $\sigma>0$ \citep[see][Ch.~13]{johnson94}, or a Laplace distributed $L\sim\textup{La}(0,b)$ with scale parameter~$b>0$ \citep[see][Ch.~24]{johnson95}. 
But the most suitable mollifier (\eg in view of simplicity of the resulting cdf) appears to be the logistic one, namely $L\sim\textup{L}(0,c)$ with scale parameter~$c>0$ \citep[see][Ch.~23]{johnson95}. Independent of the specific choice of~$L$ (although we often consider $L\sim\textup{L}(0,c)$ in the sequel), the mollified uniform distribution is defined to be the distribution of $X=U+L$, \ie its probability density function (pdf) is obtained as the convolution of the mollifier's pdf $f_L(\cdot)$ and the uniform's pdf $f_U(\cdot)$. Here, the latter equals $f_U(x)=\indfkt_{(0;1)}(x)$, where the indicator function~$\indfkt_A(x)$ equals~1 (0) if $x\in A$ ($x\not\in A$). Hence, the pdf $f(\cdot)$ of~$X$ equals
\ba
\label{MollUni_pdf}
f(x)\ =\ \int_{-\infty}^\infty f_L(x-y)\,f_U(y)\,\textup{d}y
\ =\ \int_{0}^1 f_L(x-y)\,\textup{d}y
\ =\ F_L(x)-F_L(x-1),
\ea
where $F_L(\cdot)$ denotes the mollifier's cdf. Equation \eqref{MollUni_pdf} gets particularly simple if $L\sim\textup{L}(0,c)$ with $F_L(x) = \big(1+\exp(-x/c)\big)^{-1}$, namely
\ba
\label{MollUni_logist_pdf}
f(x)
\ =\ \frac{1}{1+\exp(-\frac{x}{c})} \,\frac{1-\exp(-\frac{1}{c})}{1+\exp(\frac{x-1}{c})},
\ea
which is also plotted in Figure~\ref{figmollified}\,(a). Note that the cdf in~(b) and the quantile function (qf) in~(c) are introduced later in Section~\ref{Soft-clipping Regression Models}. 
Looking at the ``rounded edges'' for small~$c$ in Figure~\ref{figmollified}\,(a), it gets clear that also the denomination as a ``soft uniform distribution'' would be appropriate. This term, however, has been used so far for a different construction \citep[see][]{yang06}: the soft-uniform's pdf is defined in a piecewise manner (continuous, but not differentiable), where the rectangular part is expanded by a power decay on the left side and an exponential decay on the right side. At this point, it is worth mentioning that also a (centered) asymmetric distribution could be used for mollification according to \eqref{MollUni_pdf}, although the aforementioned symmetric mollifiers are probably more convenient for practice.

\smallskip
In what follows, we review and discuss some important application scenarios for the mollified uniform distribution. 
In Section~\ref{Moments and Kurtosis}, we study the mollified uniform's moment properties and show that it may serve as a model that comprises a wide range of kurtosis levels, from platykurtic over mesokurtic up to leptokurtic shapes (\ie kurtosis $<3$ or $=3$ or $>3$, respectively). It should also be pointed out that the required derivations are elementary, just making use of the convolution structure \eqref{MollUni_pdf}, such that they may be considered as a neat example for teaching. 
Section~\ref{Intermezzo: Mollified Discrete Uniform Distribution}, in turn, constitutes a brief digression to the discrete case: a discrete mollification of the discrete uniform distribution is proposed, which would again be suitable for classroom presentation. 
Section~\ref{Soft-clipping Regression Models} then moves on to the mollified uniform's cdf and qf (also recall Figure~\ref{figmollified}), and explains how the distribution might be used for expressing ``nearly linear'' generalized linear models \citep[GLMs, see][]{mcmullagh89,neuhaus11}, \eg for binary, ordinal, or bounded quantitative data and time series. 
Finally, Section~\ref{Conclusions} concludes and outlines issues for future research.
A list of abbreviations being used in this article is provided by Table~\ref{tabAbbr}.

\begin{table}[t]
\centering
\caption{Commonly occurring abbreviations in this article.}
\label{tabAbbr}

\smallskip
\begin{tabular}{lc}
\toprule
\bfseries Term & \bfseries Abbreviation \\
\midrule
Artificial neural networks & ANN \\
Autoregressive & AR \\
Clipped rectified linear unit & cReLU \\
Cumulative distribution function & cdf \\
Generalized linear model & GLM \\
Partial autocorrelation function & pacf\\
Probability density function & pdf\\
Probability mass function & pmf\\
Quantile function & qf \\
\bottomrule
\end{tabular}
\end{table}

\section{Moments and Kurtosis}
\label{Moments and Kurtosis}
The raw moments of the uniform~$U$ are easily calculated as $E[U^n]=\frac{1}{n+1}$ for $n\in\bbn_0=\{0,1,\ldots\}$. 
Since a mollified uniform random variable~$X$ equals the convolution  $X=U+L$, recall \eqref{MollUni_pdf}, its moments follow from the moments of the mollifier~$L$ by applying the binomial sum formula:
\ba
\label{MollUni_mom}
E[X^n]
\ =\ \sum_{k=0}^n \binom{n}{k}\, E[U^{n-k}]\, E[L^{k}]
\ =\ \sum_{k=0}^n \binom{n}{k}\, \frac{1}{n-k+1}\, E[L^{k}].
\ea
If using a symmetric (centered) mollifier, moments $E[L^{k}]$ of odd order~$k$ are equal to zero. Moment expressions for even order~$k$, in turn, can be found in \citet{johnson94,johnson95}. For example, $E[L^{k}] = b^k\,k!$ if $L\sim\textup{La}(0,b)$, or $E[L^{k}] = \sigma^k\,(k-1)!!$ if $L\sim\norm(0,\sigma)$, where the double factorial of the odd number~$k-1$ is the product of all odd numbers between~1 to~$k-1$. The even-order moments of the logistic mollifier $L\sim\textup{L}(0,c)$ can be generally expressed by using so-called Bernoulli numbers; in particular, $E[L^{2}] = \frac{\pi^2\, c^2}{3}$ and $E[L^{4}] = \frac{7 \pi^4\, c^4}{15}$. 

\smallskip
In the case of a symmetric mollifier~$L$, it follows that~$X$ is symmetric around its mean $E[X]=\frac{1}{2}$. So it is more appropriate to look at the central moments of~$X$ rather than at its raw moments, where obviously $E\big[(X-\frac{1}{2})^n\big] = 0$ if~$n$ is odd. For even~$n$, the central moments follow from \eqref{MollUni_mom} by again applying the binomial sum:
\ba
\label{MollUni_cmom}
E\big[(X-\tfrac{1}{2})^n\big]
\ =\ \sum_{k=0}^n \binom{n}{k}\, \frac{E[X^{k}]}{(-2)^{n-k}}.
\ea
As an example, for $n=2$, it follows together with \eqref{MollUni_mom} that $E\big[(X-\tfrac{1}{2})^2\big] = \frac{1}{12} + E[L^{2}]$, \ie the variance satisfies $V[X] = \frac{1}{12} + V[L]$, which is also implied by Bienaym\'e's identity.

\begin{figure}[t]
\center\scriptsize
	(a)\hspace{-4ex}\includegraphics[viewport=0 45 300 305, clip=, scale=0.45]{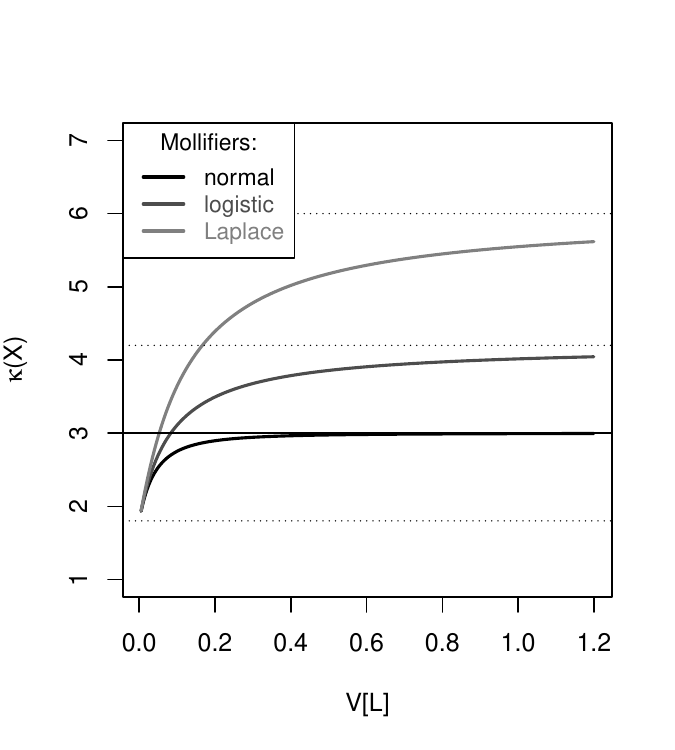}$V[L]$
	\quad
	(b)\hspace{-4ex}\includegraphics[viewport=0 45 300 305, clip=, scale=0.45]{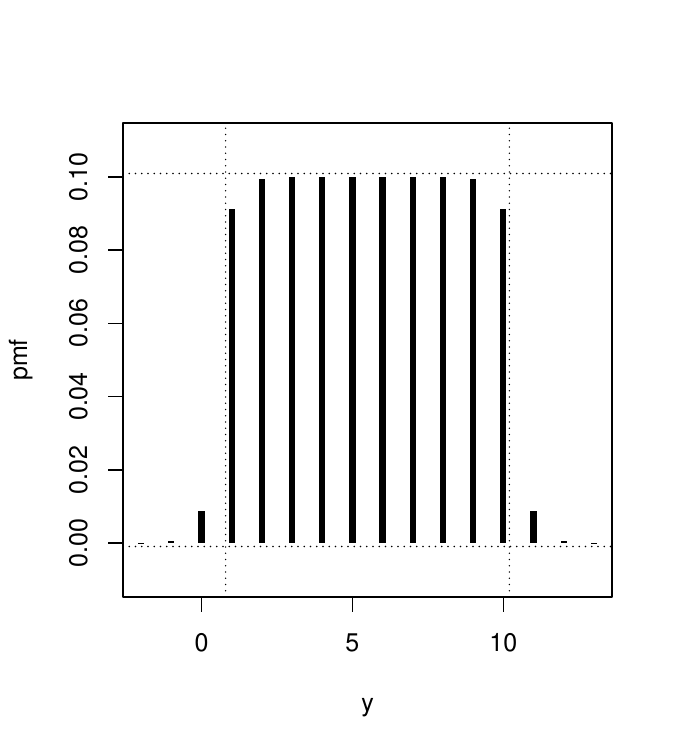}$y$
	\quad
	(c)\hspace{-4ex}\includegraphics[viewport=0 45 300 305, clip=, scale=0.45]{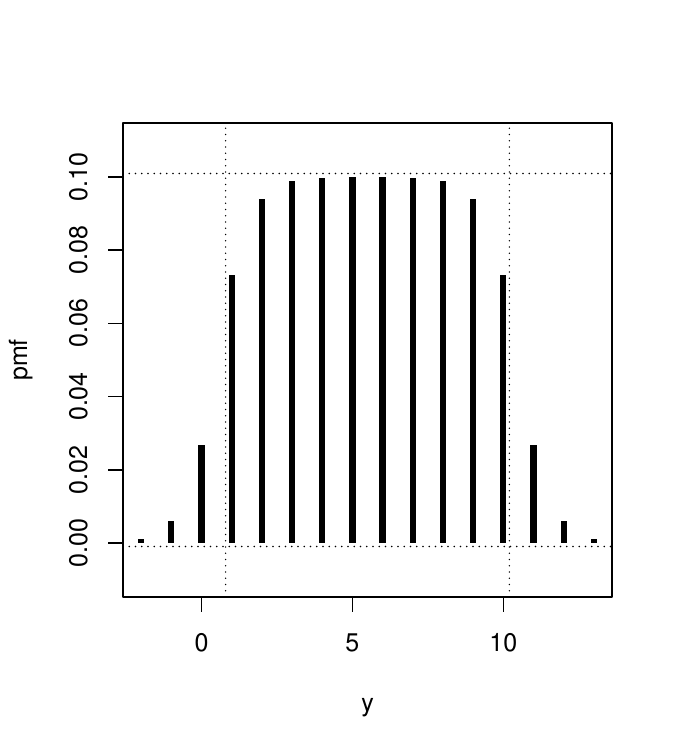}\,$y$
	\caption{Kurtosis of mollified uniform against mollifier's variance in (a). Skellam-mollified discrete uniform's pmf with $m=10$ plotted for (b) $\lambda=0.1$ and (c) $\lambda=0.5$.}
	\label{figKurtosisDiscrete}
\end{figure}

\smallskip
We shall apply formulae \eqref{MollUni_cmom} and \eqref{MollUni_mom} to compute the kurtosis of~$X$, \ie $\kappa(X) = E\big[(X-\tfrac{1}{2})^4\big] / E\big[(X-\tfrac{1}{2})^2\big]{}^2$. For a comprehensive discussion of kurtosis and possible (mis)interpretations, see \citet{westfall14}. A simple calculation leads to
$$
\textstyle
E\big[(X-\tfrac{1}{2})^4\big]
\ =\ \frac{1}{80}+\frac{1}{2}\,E[L^{2}]+E[L^{4}]
\ =\ \frac{1}{80}+\frac{1}{2}\,V[L]+V[L]^2\cdot \kappa(L),
$$
where $\kappa(L)=E[L^4]/E[L^2]^2$ is the mollifier's kurtosis (recall that $E[L]=0$). 
For the example cases discussed before, one gets
\ba
\label{kurtosis_examples}
\begin{array}{lll}
V[L]\ =\ \sigma^2
&\text{and}\quad
\kappa(L)\ =\ 3
& \text{if } L\sim\norm(0,\sigma),
\\[1ex]
V[L]\ =\ \frac{\pi^2}{3}\, c^2
&\text{and}\quad
\kappa(L)\ =\ \frac{21}{5}
& \text{if } L\sim \textup{L}(0,c),
\\[1ex]
V[L]\ =\ 2 b^2
&\text{and}\quad
\kappa(L)\ =\ 6
& \text{if } L\sim \textup{La}(0,b).
\end{array}
\ea
So finally, assuming a constant~$\kappa(L)$ like in \eqref{kurtosis_examples},
\ba
\label{MollUni_kurtosis}
\kappa(X)\ =\ \frac{\frac{1}{80}+\frac{1}{2}\,V[L]+V[L]^2\cdot \kappa(L)}{\big(\frac{1}{12} + V[L]\big)^2}
\quad \to\ \left\{\begin{array}{ll}
\frac{144}{80}=\frac{9}{5} & \text{if } V[L] \to 0,\\[1ex]
\kappa(L) & \text{if } V[L] \to \infty.
\end{array}\right.
\ea
Hence, modifying the mollifier's variance, we can move between the uniform's kurtosis~$\frac{9}{5}<3$ (platykurtic) and the mollifier's kurtosis $\kappa(L)$. 
In Figure~\ref{figKurtosisDiscrete}\,(a), the kurtosis $\kappa(X)$ is plotted against the mollifiers' variance $V[L]$ for the cases \eqref{kurtosis_examples}. The dotted lines indicate the kurtosis of $\textup{U}(0,1)$, $\textup{L}(0,c)$, and $\textup{La}(0,b)$, respectively, whereas the solid line refers to the normal's kurtosis~3. In particular, the solid line expresses the mesokurtic case, while kurtosis values below (above) the solid line correspond to the platykurtic (leptokurtic) case. If using a normal mollifier, the mollified uniform is always platykurtic, whereas for logistic or Laplace mollifiers, one can move across all three cases. Finally, recalling the pdf in Figure~\ref{figmollified}\,(a), the mollified uniform distribution also illustrates the main message in \citet{westfall14}, namely that ``kurtosis tells you very little about the peak or center of a distribution'' (p.~194).

\section{\mbox{Intermezzo: Mollified Discrete Uniform Distribution}}
\label{Intermezzo: Mollified Discrete Uniform Distribution}
Before continuing with the mollified uniform's cdf and corresponding applications in Section~\ref{Soft-clipping Regression Models}, let us have a brief look at the discrete case. In complete analogy to \eqref{MollUni_pdf}, one can define the mollified discrete uniform distribution as the convolution $Y=V+Z$, where~$V$ follows the (standard) discrete uniform distribution $\mathcal{U}\{1,m\}$ with $m\in\bbn=\{1,2,\ldots\}$ and $m\geq 2$, and where~$Z$ is symmetrically distributed around~0 with support $\bbz=\{\ldots,-1,0,1,\ldots\}$. Here, $P(V=k)=\frac{1}{m}$ for $k\in\{1,\ldots,m\}$ and~0 otherwise. 

\begin{example}
\label{exampleSkellam}
The probably most well-known candidate for~$Z$ is the Skellam distribution \citep[see][Section~4.12.3]{johnson05}, which is itself a convolution of two random variables. More precisely, $Z\sim\skellam(\lambda_1,\lambda_2)$ can be generated as the difference $Z=Y_1-Y_2$ of the two independent Poisson variates $Y_1\sim\poi(\lambda_1)$ and $Y_2\sim\poi(\lambda_2)$. Its mean and variance are given by $\lambda_1-\lambda_2 \in\bbr$ and $\lambda_1+\lambda_2> 0$, respectively. Thus, to be suitable as a mollifier, we consider the centered symmetric case $Z\sim\skellam(\lambda,\lambda)$ such that $V[Z]=2\lambda$ and $E[Z^k]=0$ for odd~$k$. 
The Skellam's probability mass function (pmf) and cdf can be computed by using pdf and cdf, respectively, of the non-central $\chi^2$-distribution, see \citet{johnson59} for details. 
\end{example}
The pmf of $Y=V+Z$ is computed in analogy to \eqref{MollUni_pdf}, namely as
\ba
\label{MollDUni_pmf}
\begin{array}{@{}rl}
P(Y=y)
\ =& \sum_{k=1}^m P(Z=y-k)\ P(V=k)
\ =\ \tfrac{1}{m}\,\sum_{k=1}^m P(Z=y-k)
\\[1ex]
=&
\tfrac{1}{m}\,\big(P(Z\leq y-1) - P(Z\leq y-m-1)\big)
\qquad\text{for } y\in\bbz.
\end{array}
\ea
The cdf follows analogously as
\ba
\label{MollDUni_cdf}
\textstyle
P(Y\leq y)
\ =\ \sum_{k=1}^m P(Z\leq y-k)\ P(V=k)
\ =\ \tfrac{1}{m}\,\sum_{k=1}^m P(Z\leq y-k)
\quad\text{for } y\in\bbz.
\ea
Illustrative plots of the pmf \eqref{MollDUni_pmf}, if using a Skellam mollifier according to Example~\ref{exampleSkellam}, are provided by parts~(b) and~(c) of Figure~\ref{figKurtosisDiscrete}. Moment calculations for the mollified discrete uniform are done in nearly the same way as in Section~\ref{Moments and Kurtosis}, as we are still concerned with a convolution of the form $Y=V+Z$. One just has to adapt the uniform's moments in \eqref{MollUni_mom}, namely to $E[V^n] = \frac{1}{m}\,\sum_{k=1}^m k^n$, which can be generally computed by using Faulhaber's formula. In particular, one gets the following (central) moments of~$V$:
$$
E[V]\ =\ \tfrac{1}{2}\,(m+1),
\quad
E\big[(V-\tfrac{m+1}{2})^2\big]\ =\ \tfrac{1}{12}\, (m^2-1),
\quad
E\big[(V-\tfrac{m+1}{2})^4\big]\ =\ \tfrac{1}{240}\, (3 m^4-10 m^2+7).
$$
The (symmetrically) mollified discrete uniform distribution has mean~$\frac{m+1}{2}$ as well, the odd central moments equal zero, whereas the even central moments follow from
\ba
\label{MollDUni_cmom}
E\big[(Y-\tfrac{m+1}{2})^n\big]
\ =\ \sum_{k=0}^n \binom{n}{k}\, E\big[(V-\tfrac{m+1}{2})^{n-k}\big]\, E[Z^{k}].
\ea
Here, we have non-zero summands only for even~$k$. For the kurtosis, we end up with
\ba
\label{MollDUni_kurtosis}
\kappa(Y)\ =\ \frac{\frac{1}{240}\,(3 m^4-10 m^2+7)+\frac{1}{2}\,(m^2-1)\, V[Z]+V[Z]^2\cdot\kappa(Z)}{\big(\frac{1}{12}\,(m^2-1)+V[Z]\big)^2}.
\ea
If using the leptokurtic Skellam mollifier having $\kappa(Z) = 3+\frac{1}{2\lambda}$, then $\kappa(Y)$ tends to the platykurtic value $\kappa(V)=\frac{3}{5}\,(3-\frac{4}{m^2-1})$ for $V[Z]\to 0$. But the mollified discrete uniform distribution may also become mesokurtic, namely if $\lambda = \frac{1}{240}\,(m^4-1)$, or even slightly leptokurtic: the maximal kurtosis is attained for $\lambda=\frac{1}{120}\, (m^4+5 m^2-6)$ and takes the value $3+\frac{30}{(m^2-1) (m^2+11)} >3$. Here, the latter result can be derived by simple calculus.

\section{Soft-clipping Regression Models}
\label{Soft-clipping Regression Models}
Let us return to the continuous mollified uniform distribution. 
The cdf~$F(x)$ of the mollified uniform distribution is generally obtained by integrating $f(x)=F_L(x)-F_L(x-1)$ according to \eqref{MollUni_pdf}. For the Laplace mollifier, however, this leads to a piecewise expression, whereas the cdf for the normal mollifier requires non-elementary functions. More precisely, let $\varphi(x)$ and $\Phi(x)$ denote the pdf and cdf of $\norm(0,1)$. Then, using that $\varphi'(x)=-x\cdot\varphi(x)$, one can show that an antiderivative of~$\Phi(x)$ is given by $\varphi(x)+x\cdot\Phi(x)$, and this allows to express the cdf in case of the normal mollifier. By contrast, quite simple expressions for cdf~$F$ and qf~$F^{-1}$ are obtained if using the logistic mollifier. Then,
\begin{align}
F(x)
\ =\ & \displaystyle
c\,\ln\left(\frac{\exp(\frac{x}{c})+1}{\exp(\frac{x-1}{c})+1}\right)
\label{softclipping}
\\
\ =\ & \min\{1,\max\{0,x\}\}\, +\, c\,\ln\!\Big(1+\exp\big(-|\tfrac{x}{c}|\big)\Big)\, -\, c\,\ln\!\Big(1+\exp\big(-|\tfrac{1-x}{c}|\big)\Big).
\nonumber
\end{align}
The second expression \citep[see][]{weissjahn24} allows for a numerically stable implementation by using the \texttt{log1p}~function, which is offered by common programming languages. In particular, it shows that $F(x)$ approaches the piecewise linear function $\textup{cReLU}(x) := \min\{1,\max\{0,x\}\}$ for $c\to 0$, also see Figure~\ref{figmollified}\,(b), which is the cdf of the uniform distribution, and which is also known as the clipped rectified linear unit (cReLU) function \citep[see][]{cai17}. So $F(x)$ constitutes a smooth version of $\textup{cReLU}(x)$ and is, thus, referred to as the soft clipping function $\textup{sc}_c(x)$ by \citet{klimek20}. The mollified uniform's qf $F^{-1}: (0;1)\to\bbr$, in turn, equals
\ba
\label{softclipping_inv}
F^{-1}(q)
\ =\ c\,\ln\biggl(\frac{\exp(\frac{q}{c})-1}{1-\exp(\frac{q-1}{c})}\biggl)
\ =\ q\, +\, c\,\ln\!\big(1-\exp(-\tfrac{q}{c})\big)\, -\, c\,\ln\!\big(1-\exp(-\tfrac{1-q}{c})\big),
\ea
see \citet{weissjahn24}, where the second expression is again suitable for the \texttt{log1p}~function. A plot of the qf is provided by Figure~\ref{figmollified}\,(c). 

\bigskip
Besides mere probabilistic computations, the cdf $\textup{sc}_c(\cdot)$ from \eqref{softclipping} and qf $\textup{sc}_c^{-1}(\cdot)$ from \eqref{softclipping_inv} can also be used for stochastic modeling. 
In the context of ANNs \citep[see]{titterington10}, it may serve as an output or activation function, which is discussed in \citet{klimek20,ohn19,jahn23} in more detail. Furthermore, \citet{weissjahn24} apply $\textup{sc}_c(\cdot)$ and $\textup{sc}_c^{-1}(\cdot)$ as response and link function, respectively, for regression analysis, namely for defining soft-clipping GLMs. More precisely, \citet{weissjahn24} proposed a time-series GLM for bounded counts having a conditional binomial distribution, while \citet{chen23} adapted this approach to a conditional discrete-beta distribution, \citet{jahnetal23} to spatio-temporal counts, and \citet{weissswidan24} to Hidden-Markov models for ordinal time series. 
Generally, soft-clipping GLMs are relevant for modeling binary, ordinal, or bounded quantitative response variables (like binomial counts or beta-distributed outcomes), see \citet{mcmullagh89,neuhaus11,tutz22} for background information. In all these cases, the logit GLM (\ie with $\textup{L}(0,1)$'s qf as the link function) and probit GLM (using $\norm(0,1)$'s qf) are most common so far. 
However, these GLMs are highly non-linear such that the interpretation of the model parameters might be difficult, see the discussion in \citet{voas24}. Therefore, the soft-clipping GLM constitutes an appealing alternative, where parameters are interpreted like for a linear model. 

\smallskip
For illustration, consider first the classical regression scenario where the response variable~$Y$ is conditionally binomially distributed depending on some explanatory variable~$x$, say $Y|x \sim\bin(n,\pi_x)$. In a GLM setup with link function~$g:(0;1)\to\bbr$ and response function $h=g^{-1}$, one assumes that $g(\pi_x)=a+b\,x$ and $\pi_x=h(a+b\,x)$, respectively. If using the logit GLM with $h(x)=\big(1+\exp(-x)\big)^{-1}$, it follows that the log-odds is linear in~$x$, namely $\ln\!\big(\pi_x/(1-\pi_x)\big) = a+b\,x$. But the relation between~$Y$ and~$x$ is highly non-linear (which also holds true for probit modeling, \ie if $h=\Phi$) and can thus cause difficulties in correctly interpreting the model, see \citet{voas24} for common misunderstandings. 
In order to achieve a linear relationship between~$Y$ and~$x$, a first idea is using the identity link (\ie assuming the exact linearity $\pi_x=a+b\,x$), which, however, requirer for strict constraints on the parameters~$a,b$ to ensure that $a+b\,x\in (0;1)$ always holds. Parameter constraints are dispensable if using the cReLU~response instead, but this choice (although often used for ANNs) has two disadvantages: cReLu is not differentiable such that common asymptotics of maximum likelihood estimation do not hold, and $\pi_x$ might become equal to~0 or~1 which leads to a degenerate binomial distribution. These problems are solved if using the soft-clipping response function $\textup{sc}_c(\cdot)$ instead, as this is infinitely differentiable and attains values only in the open interval $(0;1)$. In particular, \eqref{softclipping} is close to (piecewise) linearity provided that~$c$ is chosen sufficiently small, \eg $c=0.01$ in view of Figure~\ref{figmollified}\,(b). 

\begin{example}
\label{exampleGLM}
Let us illustrate the above discussion by a small experiment, where 1,000 data sets of length 100 are simulated as follows. First, in each simulation run, a sample of 100 explanatory variables~$x$ is generated according to $\bin(n,0.5)$ with $n=30$. Then, $\pi_x=\textup{sc}_{0.01}(a+b\,\frac{x}{n})$ with $a=0.4$ and $b=-0.6$ is computed for each~$x$, and the response~$Y$ is generated from $\bin(n,\pi_x)$. Finally, an ordinary linear model is fit to the sample of 100 pairs $(Y,x)$, \ie assuming that $E[Y]=\alpha+\beta\,x$ holds exactly. All previous steps were repeated for 1,000 times, thus leading to 1,000 estimates $(\hat{\alpha}, \hat{\beta})$.

Analyzing these 1,000 simulation runs, it gets clear that soft clipping is indeed necessary for computing~$\pi_x$, because if we would use $\pi_x=a+b\,\frac{x}{n}$ instead, then in 995 simulation runs, at least one~$\pi_x$ per sample would not be positive (in the mean, $a+b\,\frac{x}{n}\leq 0$ happens about five times per sample). Next, let us look at the estimates for the fitted linear model $E[Y]=\alpha+\beta\,x$. The sample mean across the $\hat{\alpha}$ equals $\approx 11.7$, the one across $\hat{\beta}$ equals $\approx -0.58$. Under the exact linearity $\pi_x=a+b\,\frac{x}{n}$, we would expect that $E[Y|x]=n(a+b\,\frac{x}{n}) = na+b\,x$, \ie $\alpha=12$ and $\beta=-0.6$. Looking at the aforementioned means of estimates, we recognize a close agreement, \ie the soft-clipping GLM that was used for data generation indeed behaves very similar to a truly linear model, but avoids the problem of inadmissible $\pi_x$-values. 
\end{example}
The above binomial GLM is extended to the time-series case by defining a conditional regression model like in \citet{weissjahn24}. For ease of presentation, let us focus on the most simple case of an AR$(1)$-like model (first-order autoregressive) for $(Y_t)_{t\in\bbn_0}$ with range $\{0,\ldots,n\}$ and $n\in\bbn$, where $Y_t|Y_{t-1}, Y_{t-2}, \ldots$ is assumed to be conditionally $\bin(n,\pi_t)$-distributed with $\pi_t = \textup{sc}_{c}\big(a+b\,Y_{t-1}/n\big)$. This model constitutes a stationary finite Markov chain such that all stochastic properties can be computed numerically exactly \citep[Section~3.1]{weissjahn24}. From a time series perspective, the linearity of a true AR$(1)$ model manifests itself especially in terms of the (normalized) marginal mean $E[Y_t]/n$ which we expect to be $a/(1-b)$, and of the partial autocorrelation function (pacf), which we expect to be~$b$ at lag~1 and~$0$ at lags $\geq 2$ \citep[see][]{brockwell11}. 
\begin{figure}[t]
\center\scriptsize
	(a)\hspace{-4ex}\includegraphics[viewport=0 45 405 305, clip=, scale=0.45]{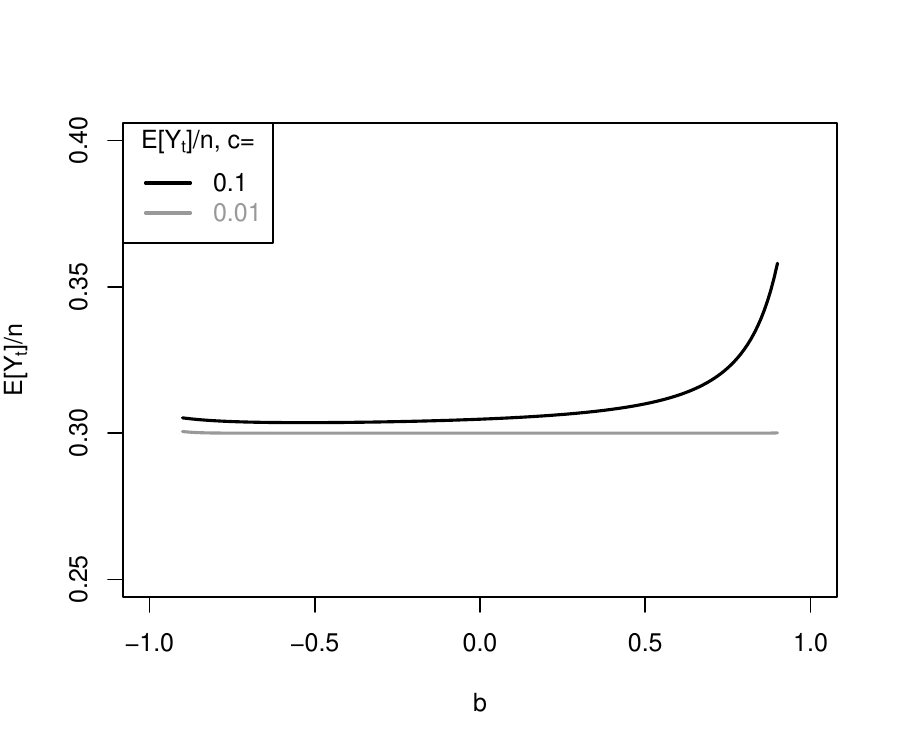}$b$
	\qquad\qquad
	(b)\hspace{-4ex}\includegraphics[viewport=0 45 405 305, clip=, scale=0.45]{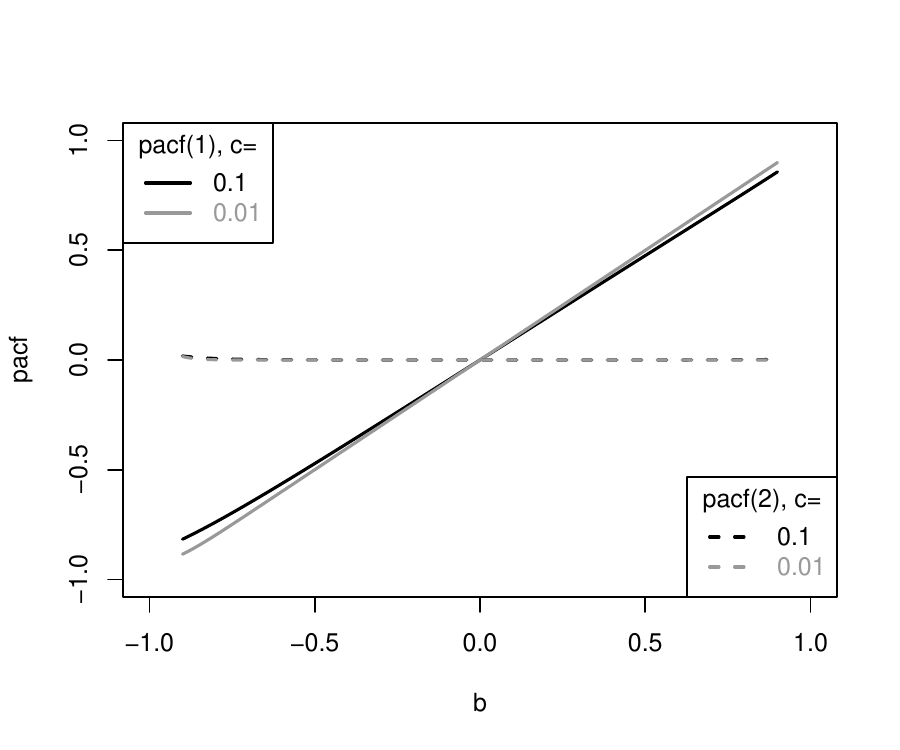}$b$
	\caption{AR$(1)$-like soft-clipping GLM with $n=30$, $a=0.3\,(1-b)$, and $-0.9\leq b\leq 0.9$: plots of (a) normalized mean and (b) pacf at lags~1 and~2.}
	\label{figscAR1}
\end{figure}
These properties are analyzed in Figure~\ref{figscAR1} for $n=30$, $a=0.3\,(1-b)$, and $-0.9\leq b\leq 0.9$, which considers the smallest and largest choice of~$c$ from Figure~\ref{figmollified}, namely $c\in\{0.01,0.1\}$. Under exact linearity, the normalized mean should be constantly equal to~$0.3$. While $c=0.1$ leads to visible deviations especially for $b\geq 0.5$, we observe a nearly perfect agreement for $c=0.01$, see Figure~\ref{figscAR1}\,(a). Analogously, the pacf values for $c=0.1$ in part~(b) deviate from the ``target values''~$b$ and~$0$, respectively, for large absolute values of~$b$, whereas we again have nearly perfect agreement for $c=0.01$. So the AR$(1)$-like soft-clipping GLM successfully approximates linearity for $c=0.01$, which is plausible also in view of Figure~\ref{figmollified}. Consequently, its model parameters are easy to interpret, namely~$a$ as the mean parameter and~$b$ as the autocorrelation parameter, and thus avoid the misunderstandings described in \citet{voas24}. In addition, the AR$(1)$-like soft-clipping GLM allows for a wide range of negative autocorrelations while being nearly linear, recall Figure~\ref{figscAR1}, and hence solves a frequently expressed criticism on many count time series models \citep[p.~13]{weiss21}. 

\smallskip
We conclude this section by briefly pointing out a further application scenario of the mollified uniform distribution, which is described in \citet{weisstestik22}. In order to investigate the effect of model misspecification (disregarded nonlinearity) in statistical process control, the soft-clipping function is used in the opposite way as before, namely for generating tailor-made deviations from linearity by choosing~$c$ sufficiently apart from zero.

\section{Conclusions}
\label{Conclusions}
In this article, the (continuous) mollified uniform distribution was rediscovered, and its origin as well as possible applications were reviewed. In addition, a novel discrete mollified uniform distribution was proposed, which uses an analogous construction as its continuous counterpart, namely a convolution of the (discrete) uniform distribution and a centered symmetric mollifier. Relevant stochastic properties can be computed by elementary calculations such that the mollified uniform distributions might also serve as neat examples in teaching. From a probabilistic point of view, the mollified uniform distribution constitutes a model that may cover a wide range of kurtosis levels. Furthermore, its cdf and qf, respectively, are used for the defining soft-clipping regression models, which can be adapted to binary, ordinal, and bounded quantitative data and time series. Here, the main feature of the soft-clipping models is given by their ability (provided that the scale parameter~$c$ is sufficiently close to zero) to nearly behave like linear models (and thus being easy to interpret) while avoiding restrictive parameter constraints and being able to handle negative dependencies. 

\smallskip
Let us conclude by outlining a couple of future research perspectives. Up to now, the soft-clipping GLM approach was only investigated in rather special setups, namely as the output function of an ANN in order to generate bounded real-valued outcomes in the unit interval $(0;1)$, see \citet{klimek20}, or for modeling bounded discrete-valued time series \citep[see][]{weissjahn24}. But applications to ordinary discrete-valued regression (without the time aspect) have not been considered so far, so soft-clipping GLMs for binary data, ordinal data (as cumulative GLMs), and count data appear to be important directions for future research. In particular, soft-clipping GLMs for bounded continuously distributed phenomena deserve attention both for traditional and time-series setups. Finally, the mollified uniform distribution could be an interesting choice for Bayesian statistics, where, similar to \citet{yang06}, it could serve as a kind of uninformative prior without hard bounds.


\subsubsection*{Conflicts of Interest}
The author declares no conflicts of interest.

\subsubsection*{Data Availability Statement}
Data sharing is not applicable to this article as no new data were created or analyzed in this study.


%
%

\end{document}